\documentclass[reprint,amsmath,amssymb,aps,prc,nofootinbib]{revtex4-1}
\usepackage{bm}% bold math
\usepackage{graphicx}
\usepackage{dcolumn}% Align table columns on decimal point
\usepackage{color}
\usepackage{epsfig, graphicx, euscript}
\newcommand{\be}{\begin{equation}}
\newcommand{\ee}{\end{equation}}
\newcommand{\bea}{\begin{eqnarray}}
\newcommand{\eea}{\end{eqnarray}}
\newcommand{\bfr}{\mbox{\boldmath $r$}}

\newcommand{\bfn}{\mbox{\boldmath $n$}}

\newcommand{\mbss}[1]{_{\mbox{\scriptsize #1}}}

\newcommand{\mbsu}[1]{\mbox{\scriptsize #1}}

\newcommand{\vphu}{\vphantom{*}}
\newcommand{\vphd}{\vphantom{1}}
\newcommand{\hfm}{\hphantom{-}}
\newcommand{\hfmm}{\hphantom{-1}}
\newcommand{\hfs}{\hphantom{^*}}

\newcommand{\ve}{\varepsilon}

\begin{document}

\title{Electric and Magnetic Moments and Transition Probabilities in $^{208}$Pb $\pm 1$ Nuclei
}

\author{V. Tselyaev}
\email{tselyaev@mail.ru}
\author{N. Lyutorovich}
\affiliation{St. Petersburg State University, St. Petersburg, 199034, Russia}
\author{J. Speth}
\affiliation{Institut f\"ur Kernphysik, Forschungszentrum J\"ulich, D-52425 J\"ulich, Germany}
\author{G. Martinez-Pinedo} \affiliation{GSI Helmholtzzentrum f\"ur
  Schwerionenforschung, Planckstra{\ss}e~1, 64291 Darmstadt, Germany}
\affiliation{Institut f{\"u}r Kernphysik (Theoriezentrum), Fachbereich
  Physik, Technische Universit{\"a}t Darmstadt,
  Schlossgartenstra{\ss}e 2, 64298 Darmstadt, Germany}
\affiliation{Helmholtz Forschungsakademie Hessen f\"ur FAIR, GSI
  Helmholtzzentrum f\"ur Schwerionenforschung, Planckstra{\ss}e~1,
  64291 Darmstadt, Germany}
\author{K. Langanke}
\affiliation{GSI
  Helmholtzzentrum f\"ur Schwerionenforschung, Planckstra{\ss}e~1,
  64291 Darmstadt, Germany}
\affiliation{Institut f{\"u}r Kernphysik
  (Theoriezentrum), Fachbereich Physik, Technische Universit{\"a}t
  Darmstadt, Schlossgartenstra{\ss}e 2, 64298 Darmstadt, Germany}
\author{P.-G. Reinhard}
\affiliation{Institut f\"ur Theoretische Physik II, Universit\"at Erlangen-N\"urnberg,
D-91058 Erlangen, Germany}
\date{\today}
\begin{abstract}
We present moments and transition probabilities in the neighboring odd-mass
nuclei of $^{208}$Pb calculated fully self-consistently from the
s.p. properties of $^{208}$Pb with polarization corrections from its
excitations, both given from previous
Skyrme-Hartree-Fock and RPA calculations. The electric
results agree nicely with the data with two very interesting
exceptions. In the magnetic case we obtain similar results. We discuss
also polarization contributions to the $l$-forbidden $M1$ transitions,
which are, however, much too small compared to the data. With a
modified external field operator which accounts effectively for
mesonic and many-body effects the description of the data can be
substantially improved.
 \end{abstract}
%\pacs{21.60.Jz}

\maketitle

\section{Introduction}
\label{sec:Intr}

Nuclear shell structure is intimately related to nuclear
single-particle (s.p.) properties as, e.g., s.p. energies with
spin-orbit splitting thereof, s.p. multipole moments, or magnetic
moments, see e.g.  \cite{DeShalit1963,Ring_1980}. A proper description
of s.p. properties had been a crucial benchmark for the development of
the empirical nuclear shell model which has become textbook standard
since long, see
e.g. \cite{DeShalit1963,BM1B,EG3B,BertschBook1972,Ring_1980}. Early
development used the properties of odd systems next to doubly magic
nuclei directly as s.p. signal. Soon it was realized that the one
nucleon added to or removed from the doubly-magic mother nucleus acts
back on the core. It responds by polarization which is determined by
low-lying collective electric and magnetic resonances of the core
nucleus.  The effect of core polarization had been taken into account
by augmenting the shell model with empirical nuclear response theory
\cite{Migdal_1967}. This then allowed reliable calculations of
s.p. moments and, more demanding, transitions strengths between
different s.p. configurations in odd nuclei \cite{Speth_1977}, the
latter being important, e.g., in astro-physical reaction chains.

The next stage in nuclear model development came up with
self-consistent models using effective interactions, better described
as nuclear density functional theory (DFT). Nearly simultaneously
appeared relativistic \cite{Duerr1956,Wal74a} and non-relativistic DFT
models \cite{Sky59a,Vau70a,Dec75a}. The advantage of self-consistent
models is that they can be extrapolated farther away from the valley
of stable nuclei than empirical models, even up to neutron stars
\cite{Horowitz_2001,ErlerTOV2013}.  Originally oriented on global
nuclear properties, as energies and radii, nuclear DFT soon has been
developed further to access more refined observables, e.g., nuclear
resonance excitations within a self-consistent
Random-Phase-Approximation (RPA) \cite{Bertsch_1975,Kre77a}.  Odd
nuclei are also naturally in reach of nuclear DFT, see e.g.
\cite{Dob01a,Borzov_2008,Borzov_2010,Tolokonnikov_2012,Saperstein_2014,Litvinova_2011,Co_2015,Sassarini_2021},
although complicated by the need of blocking and scanning a large
amount of competing configurations \cite{Pot10a}.  So far, odd nuclei
had a minority application in the world of nuclear DFT and, to
the best of our knowledge, the topic of transition strengths between
the s.p. states in odd nuclei has been addressed practically only in
the context of the non-self-consistent approach
\cite{Ring_1973a,Bauer_1973,Hamamoto_1976,Dmitriev_1983,Tselyaev_1989}
%%% has not yet been addressed in that context
(see, however, the recent paper \cite{Colo_2017}).  It is the aim of
this paper to study the description of s.p. moments and transition
strengths for odd nuclei next to $^{208}$Pb for Skyrme functionals. To
this end, we use the strategy already explored in empirical models,
namely to describe the odd nucleon (or hole) in the mean-field of the
$^{208}$Pb core and consider the self-consistent rearrangement of the
mean field perturbatively through the RPA response of the core to the
extra nucleon (or hole). This approach is legitimate in $^{208}$Pb
where one nucleon out of 208 constitutes a small perturbation, indeed.

The paper is outlined as follows: In Section~\ref{sec:theor} we
present the theoretical background which is based the many body Green
functions \cite{Migdal_1967,Speth_1977}, explain briefly the numerical
realization, and check the reproduction of low-lying resonances by the
chosen RPA scheme. In Section~\ref{sec:res} we
present our numerical results and draw various conclusions. Finally we
summarize our calculations.

\section{Formal framework}
\label{sec:theor}
\subsection{Transition operators}
\label{eq:transop}

We consider moments and transition strengths for the s.p. states of
odd-mass nuclei for electric and magnetic multipole operators ${Q}$.
Basis of the description are the s.p. states $\alpha$ of the even-even
core nucleus.  Here and in the following we label these s.p. states
briefly by numerical indices ($1,2,3,\ldots$) as synonym for
$\alpha_1,\alpha_2,\ldots$ which stand for the set of the quantum
numbers of some single-particle basis. An important aspect in
  this paper is that we consider polarization corrections
  to the measuring operator, thus dealing with an effective operator
\begin{equation}
   \tilde{Q}_{12}
  =
  {Q}_{12} +\Delta{Q}_{\mathrm{pol},12}\;.
\label{eq:qeff}
\end{equation}
%The mesonic correction
%%% $\Delta{Q}_{\mathrm{mes},12}$
%is treated here approximately, as we explain below, and considered
%only
%for the magnetic transisition where it plays a crucial role.}
\begin{figure}
\centerline{\includegraphics[width=0.7\linewidth,trim={0mm 0mm 80mm 0mm},clip]{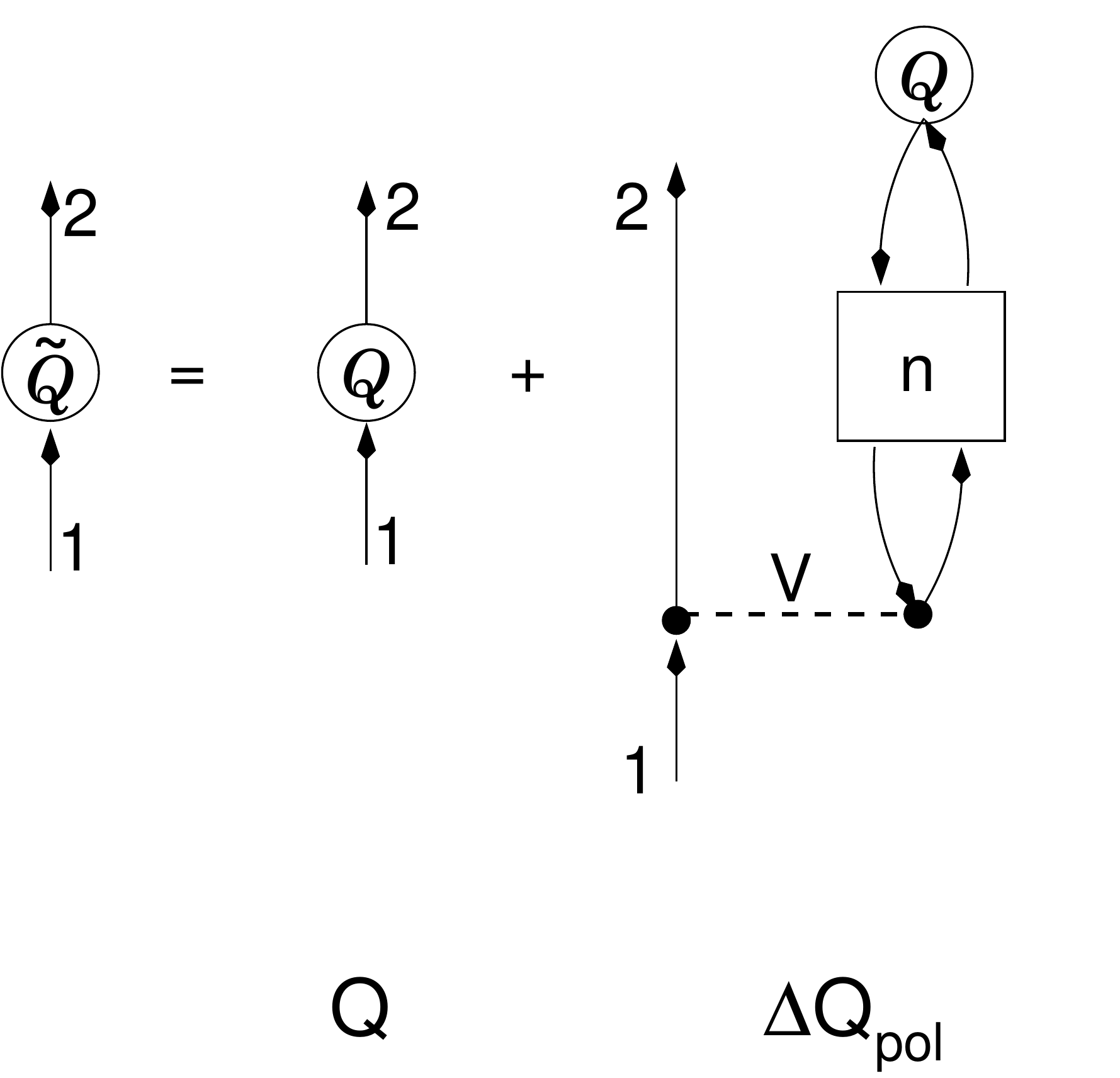}}
%\centerline{\includegraphics[width=0.8\linewidth]{diagram}}
\caption{\label{fig:diagram} Graphical representation eq.
  (\ref{eq:corepol}) for the composition of the effective operator
  $\tilde{Q}$. The term $Q$ represents the measuring operator as such,
  electric in eq. (\ref{def:qel}) or magnetic in eq. (\ref{def:qem}).
  The $\Delta Q_\mathrm{pol}$ stands for the polarization contribution
  (\ref{def:qeffpol}).}
\end{figure}
The structure of the effective measuring operator is sketched in
figure \ref{fig:diagram}. The message of the diagrams becomes clear in
the following discussion of the contributions.

The form of the bare operators
${Q}$ for the electric and magnetic moments and transitions is
given in Refs. \cite{Speth_1977,Ring_1980}.  The electric operator of
the multipolarity $L$ reads
\begin{subequations}
\label{def:qels}
\be
{Q}^\mathrm{(E)}_{\tau,LM}(\bfr) = e^{(L)}_{\tau} r^L Y_{LM}(\bfn)\,,
\label{def:qel}
\ee
where $\bfn = \bfr /r$: the $e^{(L)}_{\tau}$ are the effective charges
for the protons and neutrons which serve here to account for the
center-of-mass correction (see \cite{Eisenberg_1970}) which reads for
a nucleus with $Z$ protons and $A-Z$ neutrons
\bea
e^{(L)}_{p} &=& e \bigl[ (A-1)^L + (-1)^L (Z-1)\bigr]/A^L,
\label{def:elp}\\
e^{(L)}_{n} &=& e Z (-1/A)^L.
\label{def:eln}
\eea
\end{subequations}
In principle, the effective charges should also incorporate many-body
effects (Landau renormalization). However, in case of electric
operators Ward identities \cite{Mi_69,Speth_1977} allow to identify
the renormalized electric operator with the bare operators done
above.

The magnetic operator is defined as follows
\begin{subequations}
\label{def:qmag}
\begin{eqnarray}
{Q}^\mathrm{(M)}_{\tau,\,LM}(\bfr) &=&
\mu_{N}\,\sqrt{L\,(2L+1)}\;r^{L-1}
\nonumber\\
&\times&
%\sum_{L'=L \pm 1}
\sum_{L',\,\mu,\,\nu}(\,L',\mu,1,\nu\,|\,L,M\,)\,Y_{L',\,\mu}(\bfn)
\nonumber\\
&\times&
\biggl[\,\delta_{L',L-1}\,\biggl(\,
\tilde{\gamma}_{\tau}\,\hat{\sigma}_{1,\,\nu} +
\frac{2\xi_{\tau,\,l}}{L+1}\,\hat{l}_{\,1,\,\nu}
\biggr)
\nonumber\\
&+&
\delta_{L,1}\,\delta_{L',2}\,\frac{\kappa_{\tau}}{\sqrt{3}}\,r^2 \hat{\sigma}_{1,\,\nu}
\biggr],
\label{def:qem}
\end{eqnarray}
\begin{eqnarray}
  &&
  \tilde{\gamma}_{p\,(n)}
  =
  (1-\xi_{\,s})\,\gamma_{p\,(n)}+\xi_{\,s}\,\gamma_{n\,(p)}
  \;,
\\
  &&
  \xi_{n,\,l} = \xi_{\,l}
  \,,\quad
  \xi_{p,\,l} = 1 - \xi_{\,l}
\\
  &&
  \kappa_p = \frac{\xi_t}{100\,\mathrm{fm}^2}\;,\quad
  \kappa_n = - \frac{\xi_t}{100\,\mathrm{fm}^2}
  \;,
\label{def:kapmes}
\end{eqnarray}
where $\mu_N = e\hbar/2m_p c$ is the nuclear magneton,
$\hat{\sigma}_{1,\,\nu}$ and $\hat{l}_{\,1,\,\nu}$ are the spin Pauli
matrix and the single-particle operator of the orbital angular
momentum in the tensor representation.  The nucleons spin
gyro-magnetic moments and renormalization parameters are
\begin{eqnarray}
  &&
  \gamma_p = 2.793
  \;,\;
  \gamma_n = -1.913
  \;,\;
  \xi_s = 0.1
  \;,\;
  \xi_{\,l} = 0
\end{eqnarray}
\end{subequations}
In the case of the magnetic operator, no corresponding conservation laws
exists. Therefore, in addition to magnetic properties of bare nucleons
two effects have to be taken into account:
\\
(I) Landau renormalization \cite{Mi_69,Speth_1977} and
\\
(II) virtual exchange of mesons \cite{Chemb}.
\\
The renormalization constants $\xi_s$ and $\xi_{\,l}$ simulate the
effects of type (I).  The parameters $\kappa_{\tau}$ in the tensor
contribution $~[Y_2 \times \hat{\sigma}_1]_{1}$ simulate both type (I)
and (II) together. This term becomes important in the case of the
$l$-forbidden $M1$ transitions.
In Ref. \cite{Speth_1977} the explicit form of the effective operators is given.
Here one realizes that they
consists of a linear part which is simulated by the $\xi$-parameters
and a part with complicated many particle many hole components which
give rise to (small) vertex corrections part of which can be simulated by a term like
$[Y_2 \times \hat{\sigma}_1]_{1}$.
In the calculations we used two values: $\xi_t$ = 0 for the case where
meson effects are ignored and $\xi_t$ = 1.5 which was fitted from the
condition of describing the $l$-forbidden $M1$ transitions in the
neighboring odd-mass nuclei of $^{208}$Pb.
%which renders $\kappa_{n} = -\kappa_{p}\approx 0.0148$
%fm$^{-2}$ used in the early empirical calculations of the magnetic
%moments and transitions in Ref. \cite{Bauer_1973}.  }

The moments and the transition probabilities for the states of the odd-mass nuclei
%%% consisting of $A \pm 1$ nucleons, where $A$ is the number of the nucleons of the even-even core
are determined by the reduced matrix elements $\tilde{Q}^{L}_{(12)}$
of the multipole effective operator $\tilde{Q}^{LM}$ for which the
local external-field operator defined in
Eqs. (\ref{def:qels})--(\ref{def:qmag}) serves as the zero-order
approximation.  These reduced matrix elements are defined as
\be
\tilde{Q}^{LM}_{12} = (-1)^{j_2 - m_2}
\left(
\begin{array}{ccc}
j_1 & j_2 & L \\m_1 & -m_2 & M \\
\end{array}
\right)
\tilde{Q}^{L}_{(12)}
\label{def:rmq}
\ee
where $j$ and $m$ is the single-particle total angular moments and its projection.

For the moment $\mu^L_{(1)}$ of the multipolarity $L$ in the state with the set
of the quantum numbers $1 = \{(1),m_1\}$ and the occupation number $n^{\vphu}_{(1)}$
we have
\be
\mu^L_{(1)} = q^{\vphu}_{(1)} \sqrt{\frac{16\pi}{2L+1}}
\left(
\begin{array}{ccc}
j_1 & j_1 & L \\j_1 & -j_1 & 0 \\
\end{array}
\right) \tilde{Q}^{L}_{(11)}\,,
\label{def:moment}
\ee
where
$q^{\vphu}_{(1)} = 1 - 2 n^{\vphu}_{(1)}$ for the electric operators
and $q^{\vphu}_{(1)} = 1/2$ for the magnetic operators.
The reduced transition probability $B$ is defined through the transition
amplitude $\tilde{Q}^{L}_{(12)}$ as
\be
B(L;\,(1) \rightarrow (2)\,) =
\frac{1}{2j_1 + 1}\,\bigl(\tilde{Q}^{L}_{(12)}\bigr)^2.
\label{def:rtp}
\ee

\subsection{RPA treatment of core polarization}
\label{sec:RPA}

To lowest order approximation, the moments of an s.p. state $\alpha_1$
in the odd system are given by $Q_{11}$ and transition amplitudes as
$Q_{12}$. An important correction comes from core polarization
within the RPA, illustrated in the second term of figure
  \ref{fig:diagram}. The respective formalism was developed within
the Green-function method and described in Ref. \cite{Speth_1977}.  In
this model which is used in our present calculations, the matrix
elements of the local external-field operator ${Q}$, Eqs. (\ref{def:qels})--(\ref{def:qmag}),
are replaced by the
matrix elements of the effective (or the renormalized) operator
$\tilde{Q}$ which are determined by the solutions of the RPA
equations.  The result can be represented in the form
\cite{Ring_1973a}
\begin{subequations}
\label{eq:corepol}
\bea
\nonumber\\
  \Delta{Q}_{\mathrm{pol},12}
  &=&
  \sum_{n}\langle V|Z^{n}\rangle_{12}
\frac{\mbox{sgn}(\omega^{\vphu}_n)}{\ve^{\vphu}_{1}-\ve^{\vphu}_{2}-\omega^{\vphu}_n}
\langle Z^{n}|Q\rangle\,,
\label{def:qeffpol}
\eea
\begin{eqnarray}
  \langle Z^n|Q\rangle
  &=&
  \sum_{12} Z^{n*}_{12}\,Q^{\vphu}_{12}\,,
\label{def:zq}
\\
  \langle V|Z^{n}\rangle_{12}
  &=&
  \sum_{34} {V}^{\vphu}_{12,34} Z^{n}_{34}\,.
\label{def:g}
\end{eqnarray}
\end{subequations}
The entries of the polarization term are all quantities defined in the
even-even core: the $\omega^{\vphu}_{n}$ are the RPA excitation
energies, the $Z^{n}$ the corresponding transition amplitudes, and the
$\ve^{\vphu}_{1}$ the s.p. energies from the mean-field Hamiltonian
$\hat{h}$.  The RPA equation determining $\omega^{\vphu}_{n}$ and
$Z^{n}$ reads :
\begin{subequations}
\be
  \sum_{34} \Omega^{\mbsu{RPA}}_{12,34}\,Z^{n}_{34}
  =
  \omega^{\vphu}_{n}\,Z^{n}_{12}\,.
\label{rpaeve}
\ee
The transition amplitudes are normalized by the condition
\be
  \langle\,Z^{n}\,|\,M^{\mbss{RPA}}_{\vphd}|\,Z^{n} \rangle =
  \mbox{sgn}(\omega^{\vphu}_{n})\,,
\label{rpanorm}
\ee
where
\be
M^{\mbsu{RPA}}_{12,34} =
\delta^{\vphu}_{13}\,\rho^{\vphu}_{42} -
\rho^{\vphu}_{13}\,\delta^{\vphu}_{42}
\label{mrpa}
\ee
\end{subequations}
is the metric matrix in the RPA and $\rho$ is the single-particle density matrix
in the ground state.

Mean-field ground state and RPA are derived self-consistently from
the same given energy density functional (EDF) $E[\rho]$.
The RPA matrix $\Omega^{\mbsu{RPA}}$ is defined by
\begin{subequations}
\be
\Omega^{\mbsu{RPA}}_{12,34} =
h^{\vphu}_{13}\,\delta^{\vphu}_{42} -
\delta^{\vphu}_{13}\,h^{\vphu}_{42} + \sum_{56}
M^{\mbsu{RPA}}_{12,56}\,{V}^{\vphu}_{56,34}\,,
\label{orpa}
\ee
where the single-particle Hamiltonian $\hat{h}$ is given by the first
functional derivative of  $E[\rho]$ and
the residual interaction $\hat{V}$ by the second derivative as
\be
h^{\vphu}_{12} = \frac{\delta E[\rho]}{\delta\rho^{\vphu}_{21}}\,,
\qquad
{V}^{\vphu}_{12,34} =
\frac{\delta^2 E[\rho]}
{\delta\rho^{\vphu}_{21}\,\delta\rho^{\vphu}_{34}}\,.
\label{frpa}
\ee
\end{subequations}

The details of the solution of the equations given in this section are
the same as in the series of our previous papers, see, e.g.,
Refs. \cite{Lyutorovich_2015,Lyutorovich_2016,Tselyaev_2016,Tselyaev_2020}.
The s.p. basis was computed on a spherical coordinate-space grid with
box radius of 18~fm. The s.p. basis was limited to a maximum value of
s.p. energy as $\ve^{\mbss{max}}_{p} = 100$ MeV.

\subsection{Choice of Skyrme parametrizations
and the details of calculations}

At the side of the EDF, we use three different Skyrme
parametrizations: SLy4 \cite{CBHMS98} as an EDF with low effective
mass, SV-bas as fit to a large set of spherical nuclei and electrical
giant resonances in $^{208}$Pb (i.e.  proper core response)
\cite{Kluepfel_2009} , and SV-bas$_m$ which takes care additionally to
reproduce magnetic $M1$ response \cite{Tselyaev_2019}.

\begin{table*}[]
\caption{\label{tab:phonons}
Energies (in MeV) and the excitation probabilities $B$
of some low-lying states of $^{208}$Pb calculated within the
self-consistent RPA with for the three Skyrme-EDF parametrizations
(SV-bas, SV-bas$_{\mbsu{m}}$, and SLy4) in our survey.
Experimental data are also shown for comparison.
The $B(M1)\uparrow$ and $B(EL)\uparrow$ values are given in units of $\mu^2_N$
and $e^2$fm$^{2L}$, respectively.
}
\begin{ruledtabular}
\begin{tabular}{ccccccccccc}
&& \multicolumn{4}{c}{Energies} && \multicolumn{4}{c}{$B$} \\
 $L^{\pi}$ && SV-bas & SV-bas$_{\mbsu{m}}$ & SLy4 & Experiment
 && SV-bas & SV-bas$_{\mbsu{m}}$ & SLy4 & Experiment \\
\hline
 $1^+_2$
 && 7.95 & 7.39 & 9.67 & 7.39
 && 17.8 & 22.2 & 10.3 & 15.3 \\
 $2^+_1$
 && 4.30 & 4.29 & 4.94 & 4.09
 && 2.9$\times$10$^3$ & 2.9$\times$10$^3$ & 3.1$\times$10$^3$ & 3.2$\times$10$^3$ \\
 $3^-_1$
 && 2.97 & 3.06 & 3.48 & 2.61
 && 5.9$\times$10$^5$ & 6.3$\times$10$^5$ & 7.1$\times$10$^5$ & 6.1$\times$10$^5$ \\
 $5^-_1$
 && 3.50 & 3.89 & 4.46 & 3.20
 && 3.3$\times$10$^8$ & 3.7$\times$10$^8$ & 5.7$\times$10$^8$ & 4.5$\times$10$^8$ \\
\end{tabular}
\end{ruledtabular}
\end{table*}
As we will see, core polarization, created by the virtual excitation
of the eigenmodes in $^{208}$Pb, is crucial.  Therefore it is
important that the excitation spectrum of $^{208}$Pb is well
reproduced which is actually the case for the Skyrme parametrizations
we choose. The $E3$ and $E5$ transitions are of special interest as the
(theoretical) transition energies in the odd-mass nuclei and the
excitation energy of the $E3$ and $E5$ resonances in $^{208}$Pb can be
very similar which may give rise to resonance effects. Therefore the
single-particle spectrum of the $^{208}$Pb $\pm1$ are here of
importance. In order to demonstrate this effect we calculated those
quantities in some cases also within the Landau Migdal (LM) approach
where experimental sp energies were used and the force parameters
adjust to reproduce quantitatively excitation energies and transition
probabilities. In Table \ref{tab:phonons} the excitation energies and
transition probabilities for the first collective states of four
multipolarities are shown to give an impression. These states give
large, often dominant, contributions to the polarization effects
(which include, of course, all RPA states).

The details of the solution of the equations given in Section \ref{sec:RPA}
are the same as in the series of our previous papers, see, e.g.,
Refs. \cite{Lyutorovich_2015,Lyutorovich_2016,Tselyaev_2016,Tselyaev_2020}.
The s.p. basis was computed on a spherical coordinate-space grid with
box radius of 18~fm. The s.p. basis was limited to a maximum value of
s.p. energy as $\ve^{\mbss{max}}_{p} = 100$ MeV.

%The calculated RPA energies and transition probabilities of the
%low-lying collective states of $^{208}$Pb for all Skyrme
%parametrizations we use in the following are listed in
%Table~\ref{tab:phonons}.

%\section{Moments and transition probabilities}

\section{Results}
\label{sec:res}

\subsection{The electric case}

Here and in the following section, results for electric and magnetic
moments and transition probabilities are presented. We show only
theoretical results which can be compared with data. Before starting
the tour, we emphasize that the three Skyrme EDF are taken as
published, so to say ``from the shelves''. No re-tuning of any
parameter was done.

\begin{table}[h!]
\caption{\label{tab:E2m}
Electric quadrupole moments (in units of e$\;$fm$^2$) of the states
of the odd-mass nuclei of the lead region. The energies of the states
are listed in Table \ref{tab:Mallm}.
}
\begin{ruledtabular}
\begin{tabular}{llcccl}
 Nucleus & State & SV-bas & SV-bas$_{\mbsu{m}}$ & SLy4 & Experiment \\
\hline
 & & & & & \\
 $^{209}$Pb & $2g_{9/2}$  &    $-$25.7 &    $-$25.8 &    $-$25.6 &  $-$27(17) \cite{Chen_2015}\\
 & & & & & \\
 $^{209}$Bi & $1h_{9/2}$  &    $-$45.3 &    $-$45.5 &    $-$45.4 & $-$43.5(15) \footnotemark[1]\\
            & $1i_{13/2}$ &    $-$50.0 &    $-$50.2 &    $-$48.8 & $-$37(3) \cite{Chen_2015}\\
\end{tabular}
\end{ruledtabular}
\footnotetext[1]{Weighted mean of the values -44.6(15) \cite{Wilman_2021},
-42.0(8) \cite{Teodoro_2013}, -51.6(15) \cite{Bieron_2001},
-41.0(20) \cite{Skovpen_1981}}
\end{table}

The results for the electric quadrupole moments are shown in Table
\ref{tab:E2m}. The agreement is excellent for the both $9/2$ states
and still acceptable for the $1i_{13/2}$ state in $^{209}$Bi (last
line). The agreement is not too surprising because multipole moments,
similar as ground state deformations in even-even nuclei, are
predominantly topological quantities which are predominantly
determined by shell structure. It happens not only here but also in
level sequences that high spins still remain demanding which indicates
that mean-field models may be not yet so perfectly adjusted in that
regime.

\begin{table*}[]
\caption{\label{tab:Eallt}
Energies (in MeV) and $B(EL)$ values
for the electric multipole transitions in the odd-mass nuclei of the lead region.
}
\begin{ruledtabular}
\begin{tabular}{llcllllccccccl}
&&& \multicolumn{4}{c}{Energies} && \multicolumn{6}{c}{$B(EL)$} \\
 Nucleus & Transition && SV-bas & SV-bas$_{\mbsu{m}}$ & SLy4 & Experiment
 && SV-bas & SV-bas$_{\mbsu{m}}$ & SLy4 & LM &Experiment \\
\\
\hline
 & & & & & & & & & & & \\
\multicolumn{14}{c}{$E2$ transitions [$B(E2)$ in units of e$^2$fm$^4$]} \\
 & & & & & & & & & & & \\
 $^{207}$Tl
 &$2d_{3/2} \rightarrow 3s_{1/2}$
 && 0.774 & $\;$0.652 & 0.753 & $\hfm$0.351
 && 163 $\,$ & 161$\hfmm$ & 164$\,$ && 196(51) &\cite{Kondev_2011} \\
 & & & & & & & & & & & \\
 $^{207}$Pb
 &$2f_{5/2} \rightarrow 3p_{1/2}$
 && 0.950 & $\;$1.023 & 0.971 & $\hfm$0.570
 &&  77 $\,$ &  75$\hfmm$ &  85$\,$ && 70.9(2) &\cite{Kondev_2011} \\
 &$3p_{3/2} \rightarrow 3p_{1/2}$
 && 0.884 & $\;$0.798 & 1.103 & $\hfm$0.898
 &&  82 $\,$ &  82$\hfmm$ &  89$\,$ && 60.5(25) &\cite{Kondev_2011} \\
 & & & & & & & & & & & \\
 $^{209}$Pb
 &$4s_{1/2} \rightarrow 3d_{5/2}$
 && 0.566 & $\;$0.500 & 0.638 & $\hfm$0.465
 && 155 $\,$ & 155$\hfmm$ & 100$\,$ && 157(6) &\cite{Chen_2015} \\
 &$3d_{5/2} \rightarrow 2g_{9/2}$
 && 2.235 & $\;$2.160 & 2.519 & $\hfm$1.567
 && 245 $\,$ & 253$\hfmm$ & 236$\,$ && 184(52) &\cite{Chen_2015} \\
 & & & & & & & & & & & \\
 $^{209}$Bi
 &$2f_{7/2} \rightarrow 1h_{9/2}$
 && 1.117 & $\;$0.786 & 0.921 & $\hfm$0.896
 &&  12 $\,$ &  12$\hfmm$ &  13$\,$ && 26.1(16) &\cite{Chen_2015} \\
 &$3p_{3/2} \rightarrow 2f_{7/2}$
 && 3.042 & $\;$3.126 & 3.362 & $\hfm$2.223
 && 957 $\,$ &1100$\hfmm$ & 934$\,$ && 520(400) &\cite{Chen_2015} \\
 &$2f_{5/2} \rightarrow 1h_{9/2}$
 && 3.328 & $\;$3.281 & 3.466 & $\hfm$2.826
 && 661 $\,$ & 672$\hfmm$ & 610$\,$ && 324(44) &\cite{Chen_2015} \\
\\
 & & & & & & & & & & & & \\
\multicolumn{14}{c}{$E3$ transitions [$B(E3)$ in units of $10^3$e$^2$fm$^6$]}  \\
 $^{209}$Pb
 &$1j_{15/2} \rightarrow 2g_{9/2}$\footnotemark[1]
 && 2.001 & $\;$2.195 & 2.607 & $\hfm$1.423
 &&   133 $\,$ &  169$\hfmm$ & 257 & 58 & 67(16) &\cite{Chen_2015} \\
 & & & & & & & & & & & & \\
 $^{209}$Bi
 &$1i_{13/2} \rightarrow 1h_{9/2}$ \footnotemark[3]
 && 2.208 & $\;$1.637 & 2.359 & $\hfm$1.609
 && 25.9 $\,$ & 12.8$\hfmm$ &  23.1 & 10.2 & 15.0(15) \footnotemark[2] \\
 & & & & & & & & & & & & \\
\multicolumn{14}{c}{$E5$ transitions [$B(E5)$ in units of $10^7$e$^2$fm$^{10}$]}  \\
 $^{207}$Tl
 &$1h_{11/2}\rightarrow 2d_{3/2}$
&&  0.357 &  0.780 & 0.144 & $\hfm$0.997
&& 0.94 & 0.94 & 1.17 & 1.11 & 1.82(18) &\cite{Kondev_2011} \\
\end{tabular}
\end{ruledtabular}
\footnotetext[3]{The initial state is a mixture of
$\pi (1i_{13/2})$ and $\pi (1h_{9/2}) \otimes 3^-$.}
\footnotetext[2]{The weighted mean of $B(E3) = 1.86(22)\cdot 10^4$ \cite{Chen_2015}
and $1.2(2)\cdot 10^4$ \cite{Roberts_2016}}
\footnotetext[1]{The initial state is, a mixture of
$\nu (1j_{15/2})$ and $\nu (1h_{9/2}) \otimes 3^-$.}
\end{table*}

\begin{table}[]
\caption{\label{tab:E2mc}
Contributions (in percent) to the electric quadrupole moments and
to the transition amplitudes of the $EL$ transitions from the
external-field $EL$ operator ($Q$) and of the two RPA states of $^{208}$Pb
entering into the polarization term of Eq. (\ref{def:qeffpol}).
The electric $L^{\pi}$ RPA states with the maximum contributions are shown.
Here the $L^{\pi}_a$ is the first electric state of the respective multipolarity.
The $L^{\pi}_b$ is:
(i) the $2^+$ state from the region of the giant quadrupole resonance
with $E$ = 10.92 MeV
for the electric quadrupole moments and for the $E2$ transitions;
(ii) the $3^-$ state from the region of the giant octupole resonance
with $E$ = 5.52 MeV for the $E3$ transitions;
and (iii) the $5^-$ state with $E$ = 6.65 MeV for the $E5$ transition.
Calculations with the SV-bas parameter set.
}
\begin{ruledtabular}
\begin{tabular}{lcrcc}
 Nucleus & State/Transition & $Q$ & $L^{\pi}_a$ & $L^{\pi}_b$ \\
\hline
\multicolumn{5}{c}{Quadrupole moments}
\\
\hline
 & & & & \\
 $^{209}$Pb & $2g_{9/2}$  &   0.2 &  46.3 &  24.5 \\
 & & & & \\
 $^{209}$Bi & $1h_{9/2}$  &  56.9 &  34.6 &  15.4 \\
            & $1i_{13/2}$ &  64.7 &  29.4 &  16.2 \\
 & & & & \\
\hline
\multicolumn{5}{c}{$E2$ transitions}
\\
\hline
 $^{207}$Tl&$2d_{3/2} \rightarrow 3s_{1/2}$& 59.8 & 33.7 & 15.2 \\
 & & & & \\
 $^{207}$Pb&$2f_{5/2} \rightarrow 3p_{1/2}$&  0.2 & 48.5 & 22.4 \\
           &$3p_{3/2} \rightarrow 3p_{1/2}$&  0.2 & 45.3 & 24.7 \\
 & & & & \\
 $^{209}$Pb&$4s_{1/2} \rightarrow 3d_{5/2}$&  0.4 & 47.4 & 27.5 \\
           &$3d_{5/2} \rightarrow 2g_{9/2}$&  0.2 & 52.2 & 25.2 \\
 & & & & \\
 $^{209}$Bi&$2f_{7/2} \rightarrow 1h_{9/2}$& 51.6 & 40.7 & 16.0 \\
           &$3p_{3/2} \rightarrow 2f_{7/2}$& 42.5 & 44.5 & 16.1 \\
           &$2f_{5/2} \rightarrow 1h_{9/2}$& 36.6 & 56.7 & 13.5 \\
 & & & & \\
\hline
\multicolumn{5}{c}{$E3$ transitions}
\\
\hline
 $^{209}$Pb&$1j_{15/2} \rightarrow 2g_{9/2}$&   0.0 & 82.2 & 2.7 \\
 & & & & \\
 $^{209}$Bi&$1i_{13/2} \rightarrow 1h_{9/2}$& 12.1 & 80.9 & 2.9 \\
 & & & & \\
\hline
\multicolumn{5}{c}{$E5$ transition}
\\
\hline
 $^{207}$Tl&$1h_{11/2} \rightarrow 2d_{3/2}$&   51.9 & 13.9 & 7.3 \\
\end{tabular}
\end{ruledtabular}
\end{table}

Table \ref{tab:Eallt} collects the properties of electric multipole
transitions. We start with looking at the quadrupole case.  With
exception of the $B(E2)$ values in $^{209}$Bi, theory and experiment are
in fair agreement. The three different parameter sets give similar
results. The one case which deviates by a factor of two is the $B(E2)$
value for the "spin-flip" transition $2f_{7/2}\rightarrow 2h_{9/2}$ in
$^{209}$Bi. Their $B(E2)$ values are much smaller than those of the non
spin-flip transitions because of additional vector coupling coefficients.
For example, the corresponding non-spin flip
transition $2f_{5/2}\rightarrow 2h_{9/2}$ has (experimentally) a ten
times larger $B(E2)$ value. There is some cancellation of contributions
for the spin-flip transitions and cancellations often render the
results more volatile. Nonetheless, the two spin flip transitions in
$^{209}$Pb are in good agreement with the data and the
  qualitative difference between spin-flip and non-spin-flip
  transitions is correctly reproduced by theory. There might
still be a problem with the data where very different vales are quoted
\cite{Chen_2015} including lower values which would be much closer to
our theoretical results. The cited number in Table~\ref{tab:Eallt} is a
weighted average. But also the other two transitions which are
experimentally known are not well reproduced.

In order to understand the origin of this good agreement and the small
variation of different parameter sets we investigate composition of
ground state quadrupole moments and transitions in more detail.  Table
\ref{tab:E2mc} shows the contributions (in percentage) of the s.p
elements of the multipole operator $Q_{sp}$, of the lowest excited
${L_1}^\pi$ state, and of the giant resonance of given
multipolarity. We present the results for the SV-bas parametrization
only because the other parametrizations give similar results. For the
quadrupole moment of $^{209}$Bi one notices that the external field
operators and the polarization contributions are of the same
magnitude. We also realize that in both cases the polarization is
dominated by the lowest ${2_1}^+$ resonance and the GQR. Due to the
energy denominator in Eq. (\ref{def:qeffpol}), the contribution from the
low-lying states is of order two times larger than those from the
GQR. Similar relations are found for the quadrupole transitions
(second block in table \ref{tab:E2mc}). What changes with s.p. state
or transition is the s.p. contribution $Q_{sp}$.

%In all employed parametrizations these quantities are well
%reproduced, therefore we obtains similar results for the quadrupole
%moments as well as the B(E2) transitions the various contributions of
%which are shown in Table \ref{tab:E2tc}.

Now we look at the results for the $E3$ transition in Table
\ref{tab:Eallt}, the $1i _{13/2} \rightarrow 1h _{9/2}$ in $^{209}$Bi
and $1j _{15/2} \rightarrow 2g _{9/2}$ in $^{209}$Pb.  There is an
interesting phenomenon connected with these transitions: The
excitation energy of the collective $3^-$ resonance in $^{208}$Pb is
of the same order as the energy of the transitions. Therefore the
energy denominator in Eq. (\ref{def:qeffpol}) and the $B(E3)$ value of the
resonance plays an important role in the polarization contribution.
This too large polarization leads to a significant discrepancy between
theory and experiment. To investigate the effect we calculate the same
quantities alternatively within the empirical LM approach where, as
mentioned before, experimental s.p. energies were used and the force
parameter adjusted to reproduce quantitatively the excitation energies
and $B(EL)$ values of the lowest $E3$ and $E5$ modes in $^{208}$Pb. For this
reason, no resonance effect exists and the strong overshooting for the
$B(E3)$ values disappears. Note also that the theoretical value of the
$^{209}$Bi transition derived with the SV-bas$_{\mbsu{m}}$ parametrization
agrees nicely with the data. For that parametrization exists no
resonance
between s.p. energies and $E(3^-)$.

Finally, we look at the $B(E5)$ value of $^{207}$Tl in Table
\ref{tab:Eallt}. Here we encounter the problem that all theoretical
results including the Landau-Migdal approach are about 50 \% to small.

\subsection{The magnetic case}

In this subsection, we go for magnetic moments and transitions. We
recall from Sections \ref{eq:transop} and \ref{sec:RPA} that we employ in this case an
effective operator which simulates the effect of the Landau
renormalization as well as the
contributions of the virtual meson exchange \cite{Chemb}.

\begin{table*}[]
\caption{\label{tab:Mallm} Energies of the states (in MeV) and their
  magnetic dipole moments (in units of $\mu_N$) of the odd-mass nuclei
  next to $^{208}$Pb.
  }
\begin{ruledtabular}
\begin{tabular}{llclllcllll}
&& \multicolumn{4}{c}{Energies} && \multicolumn{4}{c}{Magnetic dipole moments} \\
 Nucleus & State & SV-bas & SV-bas$_{\mbsu{m}}$ & SLy4 & Experiment
 & $\xi_t$ & SV-bas & SV-bas$_{\mbsu{m}}$ & $\hfs$SLy4 & $\hfm$Experiment \\
\hline
 & & & & & \\
 $^{207}$Tl & $3s_{1/2}$ & $\;$0.0 & $\hfm$0.0 & 0.0 & $\hfmm$0.0
 & 0 & $\hfm$1.92 & $\hfm$2.03 & $\hfm$1.88 & $\hfm$1.876(5) \cite{Kondev_2011} \\
 & & & & &
 & 1.5 & $\hfm$1.94 & $\hfm$2.05 & $\hfm$1.90 & $\hfm$1.876(5) \cite{Kondev_2011} \\
 & & & & & \\
 $^{207}$Pb & $3p_{1/2}$ & $\;$0.0 & $\hfm$0.0 & 0.0 & $\hfmm$0.0
 & 0 & $\hfm$0.42 & $\hfm$0.44 & $\hfm$0.44 & $\hfm$0.59104(16) \footnotemark[1] \\
 & & & & &
 & 1.5 & $\hfm$0.65 & $\hfm$0.69 & $\hfm$0.68 & $\hfm$0.59104(16) \footnotemark[1] \\
 & & & & & \\
            & $2f_{5/2}$ & $\;$0.950 & $\hfm$1.023 & 0.971 & $\hfmm$0.570
 & 0 & $\hfm$0.81 & $\hfm$0.86 & $\hfm$0.86 & $\hfm$0.80(3) \cite{Kondev_2011} \\
 & & & & &
 & 1.5 & $\hfm$0.99 & $\hfm$1.06 & $\hfm$1.05 & $\hfm$0.80(3) \cite{Kondev_2011} \\
 & & & & & \\
            & $3p_{3/2}$ & $\;$0.883 & $\hfm$0.798 & 1.103 & $\hfmm$0.898
 & 0 &    $-$1.11 &    $-$1.20 &    $-$1.18 &     $-$1.09(11) \\
 & & & & &
 & 1.5 & $-$1.19 & $-$1.28 & $-$1.26 &  $-$1.09(11) \\
 & & & & & \\
            & $1i_{13/2}$ & $\;$1.584 & $\hfm$1.226 & 1.496 & $\hfmm$1.633
 & 0 &    $-$0.96 &    $-$1.00 &    $-$1.04 &     $-$1.00(3) \\
 & & & & &
 & 1.5 & $-$1.10 & $-$1.16 & $-$1.19 &  $-$1.00(3) \\
 & & & & & \\
 $^{209}$Pb & $2g_{9/2}$ & $\;$0.0 & $\hfm$0.0 & 0.0 & $\hfmm$0.0
 & 0 &    $-$1.14 &    $-$1.20 &    $-$1.18 &  $-$1.4735(16) \cite{Chen_2015} \\
 & & & & &
 & 1.5 & $-$1.28 & $-$1.36 & $-$1.35 &  $-$1.4735(16) \cite{Chen_2015} \\
 & & & & & \\
 $^{209}$Bi & $1h_{9/2}$ & $\;$0.0 & $\hfm$0.0 & 0.0 & $\hfmm$0.0
 & 0 &   $\hfm$3.42 & $\hfm$3.35 & $\hfm$3.42 & $\hfm$4.1087(5) \footnotemark[2] \\
 & & & & &
 & 1.5 & $\hfm$3.25 & $\hfm$3.19 & $\hfm$3.24 & $\hfm$4.1087(5) \footnotemark[2] \\
 & & & & & \\
            & $2f_{7/2}$ & $\;$1.117 & $\hfm$0.785 & 0.921 & $\hfmm$0.896
 & 0 & $\hfm$4.93 & $\hfm$5.04 & $\hfm$4.89 & $\hfm$4.41 \\
 & & & & &
 & 1.5 & $\hfm$5.04 & $\hfm$5.16 & $\hfm$5.01 & $\hfm$4.41 \\
 & & & & & \\
            & $1i_{13/2}$ & $\;$2.208 & $\hfm$1.637 & 2.359 & $\hfmm$1.609
 & 0 & $\hfm$7.92 & $\hfm$7.96 & $\hfm$7.78 &  $\hfm$8.07(19) \\
 & & & & &
 & 1.5 & $\hfm$8.07 & $\hfm$8.13 & $\hfm$7.94 & $\hfm$8.07(19) \\
\end{tabular}
\end{ruledtabular}
\footnotetext[1]{The weighted mean of $\mu = 0.59102(18)$ \cite{Fella_2020},
0.59064(35) \cite{Adrjan_2016}, and 0.5925(6).}
\footnotetext[2]{The weighted mean of $\mu = 4.0922(30)$ \cite{Antusek_2018},
4.0900(15) \cite{Schmidt_2018}, 4.1103(5)\cite{Chen_2015},
and 4.117(11) \cite{Feiock_1969}}
\end{table*}

Table \ref{tab:Mallm} shows the theoretical results for
magnetic dipole moments in comparison with experimental data.
The agreement is very good throughout. This happens, again,
because also he magnetic moments are dominated by ``topological'' shell
effects.

\begin{table*}[]
\caption{\label{tab:Mallt}
Energies (in MeV) and $B(ML)$ values for the magnetic transitions
in the odd-mass nuclei   next to $^{208}$Pb.
}
\begin{ruledtabular}
\begin{tabular}{llllllcccclll}
&& \multicolumn{4}{c}{Energies} && \multicolumn{6}{c}{$B(ML)$} \\
 Nucleus & Transition & SV-bas & SV-bas$_{\mbsu{m}}$ & SLy4 & Experiment
 & $\xi_t$ & SV-bas & SV-bas$_{\mbsu{m}}$ & SLy4 &&& Experiment \\
\hline
 & & & & & & & & & &&& \\
\multicolumn{13}{c}{$l$-allowed $M1$ transitions [$B(M1)$ in units of $\mu^2_N$]} \\
 & & & & & & & & & &&& \\
 $^{207}$Pb
 &$3p_{3/2} \rightarrow 3p_{1/2}$
 & 0.884 & $\;$0.798 & 1.103 & $\hfm$0.898
 & 0 & 0.36 & 0.43 & 0.41 &&& 0.45(7) \cite{Kondev_2011} \\
 &
 & & & &
 & 1.5 & 0.32 & 0.38 & 0.36
 &&& 0.45(7) \cite{Kondev_2011} \\
 & & & & & & & & & &&& \\
 &$2f_{7/2} \rightarrow 2f_{5/2}$ \footnotemark[1]
 & 2.417 & $\;$2.186 & 2.981 & $\hfm$1.770
 & 0 & 0.40 & 0.47 & 0.46 &&& 0.49(16) \cite{Kondev_2011} \\
 &
 & & & &
 & 1.5 & 0.36 & 0.42 & 0.41
 &&& 0.49(16) \cite{Kondev_2011} \\
 & & & & & & & & & &&& \\
 $^{209}$Bi
 &$2f_{5/2} \rightarrow 2f_{7/2}$
 & 2.211 & $\;$2.496 & 2.544 & $\hfm$1.930
 & 0 & 0.90 & 1.13 & 0.86 &&& 0.222(34) \cite{Chen_2015} \\
 &
 & & & &
 & 1.5 & 0.83 & 1.03 & 0.79
 &&& 0.222(34) \cite{Chen_2015} \\
 & & & & & & & & & &&& \\
\multicolumn{13}{c}{$l$-forbidden $M1$ transitions [$B(M1)$ in units of $10^{-3}\mu^2_N$]} \\
 & & & & & & & & & &&& \\
 $^{207}$Tl
 &$2d_{3/2} \rightarrow 3s_{1/2}$
 & 0.774 & $\;$0.652 & 0.753 & $\hfm$0.351
 & 0 & 2.12 & 1.12 & 2.52 &&& 23(5) \cite{Kondev_2011} \\
 &
 & & & &
 & 1.5 & 20.2 & 18.0 & 22.8
 &&& 23(5) \cite{Kondev_2011} \\
 & & & & & & & & & &&& \\
 $^{207}$Pb
 &$3p_{3/2} \rightarrow 2f_{5/2}$
% && $-$0.067 & $-$0.225 & 0.132 & $\hfm$0.328
 & $\hspace{-0.75em}-$0.067 & $\!\!\!-$0.225 & 0.132 & $\hfm$0.328
 & 0 & 1.99 & 1.36 & 1.87 &&& 50(9) \cite{Kondev_2011} \\
 &
 & & & &
 & 1.5 & 38.2 & 42.9 & 44.1
 &&& 50(9) \cite{Kondev_2011} \\
 & & & & & & & & & &&& \\
 $^{209}$Pb
 &$1i_{11/2} \rightarrow 2g_{9/2}$
 & 1.379 & $\;$1.020 & 1.554 & $\hfm$0.779
 & 0 & 0.18 & 0.08 & 0.51 &&& 9.8(11) \cite{Chen_2015} \\
 &
 & & & &
 & 1.5 & 11.6 & 14.1 & 19.2
 &&& 9.8(11) \cite{Chen_2015} \\
 & & & & & & & & & &&& \\
 $^{209}$Bi
 &$2f_{7/2} \rightarrow 1h_{9/2}$
 & 1.117 & $\;$0.786 & 0.921 & $\hfm$0.896
 & 0 & 1.04 & 0.46 & 1.12 &&& 4.6(9) \cite{Chen_2015} \\
 &
 & & & &
 & 1.5 & 14.6 & 13.3 & 18.0
 &&& 4.6(9) \cite{Chen_2015} \\
 & & & & & & & & & &&& \\
 \hline
\multicolumn{13}{c}{$M2$ transitions [$B(M2)$ in units of $\mu^2_N\;$fm$^2$]} \\
$^{209}$Pb &$1j_{15/2}\rightarrow 1i_{11/2}$
& 0.623 & 1.175 &  1.052 & $\hfm$0.644
&&  33.1 &  44.7 &  60.9  &&& 33(8) \cite{Chen_2015}\\
$^{209}$Bi &$1i_{13/2}\rightarrow 1h_{9/2}$ \footnotemark[2]
& 2.208 & 1.637 &  2.359 & $\hfm$1.609
&& 110.1 & 140.2 & 64.2  &&& 34(5)\footnotemark[3]\\
 & & & & & & & & & &&& \\
\multicolumn{13}{c}{$M4$ transitions [$B(M4)$ in units of $10^5\mu^2_N\;$fm$^6$]} \\
 $^{207}$Tl &$1h_{11/2}\rightarrow 2d_{3/2}$
&  0.357 &  0.780 & 0.144 & $\hfm$0.997
&&  4.91  &  5.67  & 4.31  &&& 2.39(23) \cite{Kondev_2011}\\
\end{tabular}
\end{ruledtabular}
\footnotetext[1]{The initial state, $E_i = 2339.921$, is a mixture of
$\nu(2f_{7/2})$ and $\nu(1i_{13/2})\otimes 3^-$ }
\footnotetext[2]{The initial state is a mixture of
$\pi (1i_{13/2})$ and $\pi (1h_{9/2}) \otimes 3^-$.}
\footnotetext[3]{Weighted mean of 18.6(104) \cite{Chen_2015}
and 38(5) \cite{Roberts_2016}}
\end{table*}

\begin{table}[h!]
\caption{\label{tab:Mallmc}
Contributions (in percent) to the magnetic dipole moments and to the
transition amplitudes of the $ML$ transitions from the
external-field $ML$ operator ($Q$) and of the two RPA states of $^{208}$Pb
entering into the polarization term of Eq. (\ref{def:qeffpol}).
The magnetic $L^{\pi}$ RPA states with the maximum contributions are shown.
Here the $L^{\pi}_a$ is:
the $1^+_2$ state with $E$ = 7.95 MeV
(representing the isovector $M1$ resonance)
for the magnetic dipole moments and for the $l$-allowed $M1$ transitions;
the $2^-$ state from the region of the giant $M2$ resonance
with $E$ = 8.56 MeV for the $M2$ transitions;
the $4^-$ state from the region of the giant $M4$ resonance
with $E$ = 7.97 MeV for the $M4$ transition.
The $L^{\pi}_b$ is:
the $1^+_1$ state with $E$ = 5.66 MeV
for the magnetic dipole moments and for the $l$-allowed $M1$ transitions;
the $2^-$ state from the region of the giant $M2$ resonance
with $E$ = 9.39 MeV for the $M2$ transitions;
the $4^-$ state from the region of the giant $M4$ resonance
with $E$ = 5.19 MeV for the $M4$ transition.
Calculations with the SV-bas parameter set and $\xi_t$ = 0.
}
\begin{ruledtabular}
\begin{tabular}{lcrrr}
%%% Nucleus & State/Transition & $Q\hfm$ & $1^+_2\hfm$ & $1^+_1\hfm$ \\
 Nucleus & State/Transition & $Q\hfm$ & $L^{\pi}_a\hfm$ & $L^{\pi}_b\hfm$ \\
\hline
\multicolumn{5}{c}{Magnetic dipole moments} \\
\hline
 $^{207}$Tl & $3s_{1/2}$  & $\hfm$120.9$\hfs$ &  $-$12.4$\hfs$ &   $-$7.8$\hfs$ \\
%%%            & $3s_{1/2}$  & $\hfm$120.9$^*$ &    $-$12.4$^*$ &     $-$7.8$^*$ \\
 & & & & \\
 $^{207}$Pb & $3p_{1/2}$  & $\hfm$114.0$\hfs$ &  $-$11.4$\hfs$ &  $-$3.4$\hfs$ \\
%%%            & $3p_{1/2}$  & $\hfm$108.9$^*$ &    $-$7.2$^*$ &     $-$2.2$^*$ \\
            & $2f_{5/2}$  & $\hfm$127.8$\hfs$ &  $-$26.1$\hfs$ &     $-$1.4$\hfs$ \\
            & $3p_{3/2}$  & $\hfm$129.4$\hfs$ &  $-$25.7$\hfs$ &     $-$3.8$\hfs$ \\
            & $1i_{13/2}$ & $\hfm$150.1$\hfs$ &  $-$34.0$\hfs$ &    $-$15.7$\hfs$ \\
 & & & & \\
 $^{209}$Pb & $2g_{9/2}$  & $\hfm$126.9$\hfs$ &    $-$25.7$\hfs$ &     $-$0.6$\hfs$ \\
 & & & & \\
 $^{209}$Bi & $1h_{9/2}$  &  $\hfm$87.9$\hfs$ &  $\hfm$9.0$\hfs$ &  $\hfm$3.0$\hfs$ \\
            & $2f_{7/2}$  & $\hfm$108.0$\hfs$ &     $-$4.8$\hfs$ &  $-$3.1$\hfs$ \\
            & $1i_{13/2}$ & $\hfm$105.1$\hfs$ &     $-$4.3$\hfs$ &  $-$0.7$\hfs$ \\
 & & & & \\
\hline
\multicolumn{5}{c}{$l$-allowed $M1$ transitions} \\
\hline
%%% & & & & \\
 $^{207}$Pb&$3p_{3/2} \rightarrow 3p_{1/2}$
 & 135.4$\hfs$ &  $-$30.5$\hfs$ &   $-$4.4$\hfs$ \\
           &$2f_{7/2} \rightarrow 2f_{5/2}$
 & 145.5$\hfs$ &  $-$40.3$\hfs$ &   $-$3.6$\hfs$ \\
 & & & & \\
 $^{209}$Bi&$2f_{5/2} \rightarrow 2f_{7/2}$
 & 141.2$\hfs$ &  $-$23.0$\hfs$ &  $-$17.0$\hfs$ \\
 & & & & \\
\hline
\multicolumn{5}{c}{$M2$ transitions} \\
\hline
%%% & & & & \\
 $^{209}$Pb&$1j_{15/2} \rightarrow 1i_{11/2}$
 &   223.4$\hfs$ & $-$27.9$\hfs$ & $-$8.4$\hfs$ \\
 & & & & \\
 $^{209}$Bi&$1i_{13/2} \rightarrow 1h_{9/2}$
 &    160.0$\hfs$ & $-$18.1$\hfs$ & $-$4.3$\hfs$ \\
 & & & & \\
\hline
\multicolumn{5}{c}{$M4$ transition} \\
\hline
 $^{207}$Tl&$1h_{11/2} \rightarrow 2d_{3/2}$
 &   147.4$\hfs$ & $-$7.1$\hfs$ & $-$4.5$\hfs$ \\
\end{tabular}
\end{ruledtabular}
\end{table}

\begin{table}[h!]
\caption{\label{tab:M1f}
The same as in Table \ref{tab:Mallmc} but for the $l$-forbidden $M1$ transitions
with two variants of the choice of the parameter $\xi_t$.
The $1^+_a$ is the $1^+_2$ RPA state of $^{208}$Pb for all the transitions.
The $1^+_b$ is the $1^+_1$ RPA state for all the transitions
except for the transition $3p_{3/2} \rightarrow 2f_{5/2}$ in $^{207}$Pb.
For the latter transition, $1^+_b$ is the high-energy $1^+$ RPA state with
$E$ = 17.71 MeV in the case of $\xi_t$ = 0 and with
$E$ = 28.95 MeV in the case of $\xi_t$ = 1.5.
}
\begin{ruledtabular}
\begin{tabular}{llcrrr}
 Nucleus & Transition  & $\xi_t$ & $Q\hfm$ & $1^+_a$ & $1^+_b$ \\
\hline
 & & & & \\
 $^{207}$Tl&$2d_{3/2} \rightarrow 3s_{1/2}$
 & 0   & 0.0  & $\hfm$54.3 & $\hfm$39.1 \\
           &
 & 1.5 & 99.6 & $\hfm$16.8 & $\hfm$12.3 \\
 & & & & \\
 $^{207}$Pb&$3p_{3/2} \rightarrow 2f_{5/2}$
 & 0   & 0.0  & $\hfm$92.9 &  $-$6.3 \\
           &
 & 1.5 & 110.8& $\hfm$20.2 &  $-$4.2 \\
 & & & & \\
 $^{209}$Pb&$1i_{11/2} \rightarrow 2g_{9/2}$
 & 0   & 0.0  &$\hfm$140.8 &  $-$46.2 \\
           &
 & 1.5 & 123.2& $\hfm$16.6 &  $-$5.5 \\
 & & & & \\
 $^{209}$Bi&$2f_{7/2} \rightarrow 1h_{9/2}$
 & 0   & 0.0  & $\hfm$58.8 & $\hfm$37.4 \\
           &
 & 1.5 & 106.6& $\hfm$15.0 & $\hfm$9.7 \\
\end{tabular}
\end{ruledtabular}
\end{table}

Table \ref{tab:Mallt} shows results for magnetic transitions. We look
first at dipole transitions.  The upper part of the dipole block sows
the allowed $M1$ transitions. Theory and experiment are in fair
agreement for $^{207}$Pb and for all three parameter sets with
variations within the three parametrizations between 10 $\%$ and
20$\%$. As we found already for electrical transitions, agreement with
data is poor for $^{209}$Bi which indicates, again, that present EDFs
still have weak points concerning high-spin s.p. states.  Below the
$l$-allowed $M1$ transitions follow the $l$-forbidden $M1$ transitions.
The external-field operator $Q$, Eq. (\ref{def:qem}), gives no
contribution to lowest order, i.e. if we neglect the $~[Y_2\times
  \hat{\sigma}]_{1}$ term ($\xi_t$ = 0). In that case, only
polarization effects yield finite contributions. However, this
contribution is one order of magnitude too small, except for
$^{209}$Bi where polarization amounts to nearly 25$\%$ of the measured
$B(M1)$.  Here is the place where the corrections through the tensor
term $\sim [Y_2\times \hat{\sigma}]_{1}$ in the magnetic operator
(\ref{def:qmag}) may become qualitatively important.  We activate the
term by setting $\xi_t=1.5$ which improves the agreement with data
dramatically. This encouraging result calls for for further analysis
of the meson-exchange currents \cite{Chemb} in the measurement of
magnetic observables.

The dominant relative contributions to magnetic moments and
transitions are collected in Table \ref{tab:Mallmc}.  We show (in
percent) the contribution from the external-field $ML$ operator
${Q}$ and the first and second $L^\pi$ magnetic RPA states in
$^{208}$Pb calculated with the SV-bas parameter set.  For $L=1$ these
two states are isoscalar $1^+_1$ with the calculated $E(1^+_1)$ = 5.66
MeV and $B(M1;\,1^+_1)$ = 5.6 $\mu_N^2$ and the isovector $1^+_2$
resonance with $E(1^+_2)$ = 7.95 MeV and $B(M1;\,1^+_2)$ = 17.8
$\mu_N^2$.  The contribution of the external-field $M1$ operator
dominates in all moments and $l$-allowed transitions. Totally different
look the $l$-forbidden $M1$ transitions where
in the case $\xi_t$ = 0
the $M1$ operator, by definition, allows no transitions.
In the case $\xi_t$ = 1.5,
the contribution of the first term $Q$
in the right-hand side of Eq. (\ref{eq:qeff}) dominates.
As can be seen from Table \ref{tab:Mallt},
the choice $\xi_t$ = 1.5 provides on the whole
the reasonable description of the $l$-forbidden $M1$ transitions,
though for the transition $2f_{7/2} \rightarrow 1h_{9/2}$ in $^{209}$Bi
one obtains the exceeding of the $B(M1)$ in about 3 times.
The contribution of the $1^+_2$
state is larger than the $1^+_1$ state as the $B(M1;\,1^+_2)$ value is
larger than the $B(M1;\,1^+_1)$ by more than a factor of three.

Finally we look at the results for $M2$ and $M4$ transitions in the lower
part of Table \ref{tab:Mallt}. The agreement with the data is not as
satisfying as for the allowed $M1$ transitions. One observers also
larger differences between the parameter sets, e.g. the result of the
SV-bas set for $^{209}$Pb agrees with the experimental value but is a
factor three to large for $^{209}$Bi.  It is worthwhile to counter
check the contributions to the total $M2$ and $M4$ transition amplitudes
in Table \ref{tab:Mallmc}.  This reveals that the external field
operator for $M2$ transitions alone produce results which are one order
of magnitude larger than the experiment value.  The polarization
contributions correct that overestimation toward the data.  One should
not wonder that the numbers do not add up to 100\%.  What is missing
is the accumulated further reduction by all other magnetic modes in the
spectrum which is in that case obviously a large fraction.  In the
case of the known $M4$ transition, the external field contribution
alone is five times larger than the experimental value $B(M4)$.  Core
polarization helps a large way to reduce the theoretical $B(ML)$ values
thus reducing the disagreement between the external field results and
the data appreciably, though not yet completely.

\section{Conclusion}

We present results of self-consistent calculations for electric and
magnetic moments and transition probabilities in the neighboring odd-mass
nuclei of $^{208}$Pb.  Starting point are Skyrme-Hartree-Fock
calculations for the ground state of $^{208}$Pb and subsequent
Skyrme-RPA to obtain the excitation spectrum of $^{208}$Pb.  From
that, we deduce the electric and magnetic s.p. matrix elements of the
odd system from those of $^{208}$Pb together with a polarization
correction to account for the change of the mean field by the odd
nucleon (hole).  For the calculations, we use a Skyrme energy
functional with parameter sets which had been previously optimized for
nuclear structure properties. We obtain at once theoretical results
which are in fair agreement with the data for the moments and the
electric as well as allowed magnetic transitions.

We also investigated the impact of the various contributions to the
final results. In the electric case for odd-neutron
%(hole)
neighbors, the whole effect comes from the polarization as the external field
gives zero contributions (except for the very small center of mass
corrections).  For odd-proton neighbors, the polarization contributes
roughly $50\% $ to moments and transitions.  In all cases, the low
lying $2^+$ state of $^{208}$Pb contributes most because of the energy
denominator in the response function. In the magnetic case, one has to
distinguish between the $M1$ properties and the higher ML
transitions. For M1, the core polarization is much smaller than in the
electric case as only the two spin orbit partners contribute (as
isoscalar and isovector state, respectively). For the higher $L$
values the polarization effects are large as many components
contribute. We obtain very similar results for three different
parameter sets which is no surprise because all three sets produce
similar excitations in $^{208}$Pb.

A particular case are $l$-forbidden $M1$ transitions. Polarization
effects produce a finite contribution which, however, is an order of
magnitude too small to reproduce the data. It is here where mesonic
contributions and vertex corrections dominate. We have simulated them
by the
%% a formerly
empirically tuned tensor term and find that
%% confirmed also
it considerably improves the agreement with data
in our self-consistent context. This preliminary result points toward
the next task, a fully microscopic description of mesonic effects and
vertex correction in magnetic transitions.

%\newpage

\begin{acknowledgements}
This work was supported by the Russian Foundation for Basic Research,
project number 21-52-12035, and the Deutsche Forschungsgemeinschaft,
contract RE 322/15-1.  This research was carried out using
computational resources provided by the Computer Center of
St. Petersburg State University.
\end{acknowledgements}

\bibliographystyle{apsrev4-1}
\bibliography{TTT}

%\begin{thebibliography}{99}
%
%
%\end{thebibliography}
%
\end{document}